\documentclass[11pt]{article}

\usepackage{amssymb,epsfig,fullpage}

\newtheorem{thm}{Theorem}{\bfseries}{\itshape}

\begin{document}
	
\title{Dynamics of multi-stage infections on networks}

\author{N. Sherborne, \hspace{0.5cm}K.B. Blyuss\thanks{Corresponding author. Email: k.blyuss@sussex.ac.uk}, \hspace{0.5cm}I.Z. Kiss 
\\\\ Department of Mathematics, University of Sussex, Falmer,\\
Brighton, BN1 9QH, United Kingdom}

\maketitle
	
\begin{abstract}
This paper investigates the dynamics of infectious diseases with a non-exponentially distributed infectious period. This is achieved by considering a multi-stage infection model on networks. Using pairwise approximation with a standard closure, a number of important characteristics of disease dynamics are derived analytically, including the final size of an epidemic and a threshold for epidemic outbreaks, and it is shown how these quantities depend on disease characteristics, as well as the number of disease stages. Stochastic simulations of dynamics on networks are performed and compared to the results of pairwise models for several realistic examples of infectious diseases to illustrate the role played by the number of stages in the disease dynamics. These results show that a higher number of disease stages results in faster epidemic outbreaks with a higher peak prevalence and a larger final size of the epidemic. The agreement between the pairwise and simulation methods is excellent in the cases we consider.
\end{abstract}

\section{Introduction}

	Mathematical models of infectious diseases are known to provide an invaluable insight into the mechanisms driving disease invasion and spread. In many cases, to obtain the first approximation of the spread of a disease it is sufficient to use a version of the classical SIR model (Kermack and McKendrick 1927) \cite{SIR}. However, major outbreaks of avian and swine influenza (Ferguson et al. 2006) \cite{fer06}, SARS (Donnelly et al. 2003) \cite{sars_paper}, and more recently, ebola (Chowell and Nishiura 2014) \cite{CN14}, have highlighted the need for a more accurate description of the disease dynamics that would provide predictive power to be used for developing measures for disease control and prevention (Keeling and Rohani 2008) \cite{keeling2008modeling}. 
	
	One of the major simplifying assumptions often used in mathematical models of disease dynamics is the exponential distribution of infectious periods. Effectively, this means that the chance of an individual recovering during any given time period does not depend on the duration of time that individual has already been infected. Whilst such an assumption may provide significant mathematical convenience and be reasonably realistic in some situations, most often it is violated, and this requires the inclusion of the precise distribution of infectious periods in the model (Bailey 1954; Hope-Simpson 1952) \cite{Bailey,simpson1952infectiousness}. There are several methods that can be employed to explicitly include a non-exponential distribution, including a multi-stage approach (Anderson and May 1992; Cox and Miller 1965) \cite{anderson1980spread,CM65}, an integro-differential formulation (Kermack and McKendrick 1927; Hethcote and Tudor 1980; Keeling and Grenfell 1997) \cite{SIR,Heth1980,keel1997gren}, and a PDE-based formulation akin to that for age-structured models (Anderson and May 1992) \cite{AM92}. In the multi-stage framework, it is assumed that the infectious stage of a disease is characterised by a number $K$ of distinct stages  (Cox and Miller 1965; Lloyd 2000; Lloyd 2001) \cite{CM65,infectperiod,lloyd2001destabilization}, with the duration of each stage being an independent exponentially distributed random number. Due to the fact that the sum of independent exponentially distributed random variables obeys a gamma distribution (Durrett 2010) \cite{durrett2010probability}, one can replace an exponential distribution with the mean infectious period $1/\gamma$ by a gamma distribution $\Gamma(K,1/(K\gamma))$ that has the same mean infectious period $1/\gamma$. The so-called {\it linear chain trick}  (Cox and Miller 1965; MacDonald 1978) \cite{CM65,Macd77} then consists in replacing a single infectious stage with $K$ identical exponentially distributed sub-stages, each having a mean period $1/(K\gamma)$. These multiple stages of infection can be used to represent periods of increased or decreased risk of transmitting the disease (Ma and Earn 2006) \cite{ma2006generality}. The same approach can be extended to models with multiple classes (Keeling and Grenfell 2002; Nguyen and Rojani 2008) \cite{keeling2002understanding,nguyen2008noise}, as well as non-exponentially distributed latency and temporary immunity periods (Blyuss and Kyrychko 2010; Wearing and Rojani 2005) \cite{BK10,wearing2005appropriate}. Following the methodology of introducing multi-stage of infection to better represent the distribution of infectious periods, we proceed with dividing the infected population into $K$ identical stages $I_1, I_2, \ldots, I_K$ to create the so-called $SI^KR$ model (Lloyd 2000) \cite{infectperiod}, and we denote the total infected population by $I = \sum_{i=1}^K I_i$. One should note that $K\gamma$ is now used as the transition rate between successive infectious stages in order to keep the average duration of infection as $1/\gamma$. With these notations, the $SI^KR$ model takes the form
	\begin{equation}\label{eq:stage_ode}
	\begin{array}{l}
	dS/dt = -\beta S I, \\
	dI_1/dt = \beta S I - K\gamma I_1, \\
	dI_2/dt = K\gamma I_1 - K\gamma I_2, \\
	\vdots \\
	dI_K/dt = K\gamma I_{K-1} - K\gamma I_K,\\
	dR/dt = K\gamma I_K,
	\end{array}
	\end{equation}
	where $S$ denotes the proportion of susceptible individuals, $R$ is the proportion of recovered or removed individuals, $\beta$ is the disease transmission rate taken to be the same for all stages of infection, and the disease is assumed to confer a life-long immunity. The importance of including not just the mean infectious period, but the actual distribution of infectious periods, as achieved by the system (\ref{eq:stage_ode}) is further highlighted by the inspection of actual values of epidemiological parameters for several real diseases as presented in Table~\ref{tab:distributions}. This table illustrates that whilst the transmission rate and the average infectious period may vary between different diseases, in all of these cases the number of stages that has to be included in order to correctly represent the disease dynamics may also be quite high, this reinforces an earlier observation about the non-exponential nature of infectious period distribution.
		
	Whilst this method of introducing multiple stages of infection is clearly more realistic, the assumption of a homogeneous fully mixed population remains very important, having significant effects on the disease dynamics (Keeling and Rohani 2008) \cite{keeling2008modeling}. Although this assumption often provides a good approximation that helps reduce complexity of the model, in many cases it is just not realistic and results in erroneous conclusions about the onset and development of epidemic outbreaks (Bansal et al. 2007; Burr and Chowell 2008) \cite{Bansal,Burr}. To address this issue, networks have been and are being used successfully to model the contact structure of the population to a high degree of detail (Danon et al. 2011; Keeling and Eames 2005) \cite{danon,keeling_eames}. Typically, network models are parameterised with empirical data or synthetic models that can be either purely theoretical, e.g. homogeneous random networks or Erd{\H o}s-R{\'e}nyi random graphs, or obey some widely observed network characteristics, such as a particular degree distribution or clustering. However, with added model realism comes complexity, which in the case of epidemic network models can be handled via mean-field models, such as pairwise models (Keeling 1999; House and Keeling 2011a) \cite{keeling1999,house2011insights} that are able to better account for the explicit nature of network links. As long as such mean-field models provide a good approximation to the explicit stochastic network models, they open up the possibility to analyticaly compute important quantities such as epidemic threshold, final epidemic size and so on. Thus, the explicit stochastic network simulation model and the pairwise model combine favourably to provide a more accurate model with some degree of analytical tractability.
	
	In this paper we are concurrently relaxing the assumptions of homogeneous random mixing and exponentially distributed infectious periods to generate a multi-stage pairwise model for the spread of epidemics on networks. The paper is organised as follows. The next Section contains a brief summary and discussion of earlier results on the properties of the $SI^KR$ model (\ref{eq:stage_ode}). In Section 3 we employ the framework of pairwise approximations to derive a multi-stage infection pairwise model and use this to derive analytical expressions for the probability of transmission of infection along an infected edge in a network, a threshold parameter controlling the onset of epidemic outbreaks, and the final size of an epidemic. In Section 4 numerical simulations of the pairwise and the full network models are performed using realistic parameter values from Table~\ref{tab:distributions} to investigate the accuracy of pairwise approximation and to illustrate the role played by the number of stages in the multi-stage distribution in the disease dynamics. The paper concludes in Section 5 with discussion of results and future outlook.
	
	\begin{table}
		\caption{Estimates of epidemiological parameters for different infectious diseases.}
		\label{tab:distributions}
		\begin{center}
			\begin{tabular}{c|cccl}
				Disease 		& $\beta$	& $\gamma^{-1}$ (days) & Stages $K$ & Source(s) \\
				\hline
				Measles	& Seasonal & 5 & 20 &	(Hope-Simpson 1952) \cite{simpson1952infectiousness}\\
				SARS  & 0.545 & 5-6 & 3	& (Bauch et al. 2005; Riley et al. 2003) \cite{bauch2005dynamically, riley2003transmission}		\\
				Influenza	& 1.66 & 2.2  & 3 & (Keeling and Rojani 2008; Cauchemez et al. 2004) \cite{keeling2008modeling, cauchemez2004bayesian} \\
				Smallpox	& 0.49 & 8.6 & 4 & (Ferguson et al. 2003; Koplan, Azizullah and Foster 1978) \cite{ferguson2003planning, koplan1978urban}
			\end{tabular}
		\end{center}
	\end{table}
	
	\section{Dynamics of the well-mixed model}
	
	As a first step, we consider the $SI^KR$ model (\ref{eq:stage_ode}), which has an implicit assumption that every member of the population has a sufficient level of contact so that the infection can be passed from any individual to any other. This is a natural extension of the basic SIR model (Kermack and McKendrick 1927) \cite{SIR}, and as such it has been well-studied in a number of papers (Lloyd 2000; Ma and Earn 2006; van den Driessche and Watmough 2002)\cite{infectperiod,ma2006generality,van2002reproduction}.
	
	Perhaps, one of the most important and commonly used parameters characterising the severity of epidemics and stability of the disease-free equilibrium is the {\it basic reproduction number} $\mathcal{R}_0$ defined as the expected number of secondary infections caused by a single typical infectious individual in a wholly susceptible population. The value of $\mathcal{R}_0$ is related to the stability of the disease-free equilibrium, and it is an important threshold parameter signifying that an epidemic will spread when $\mathcal{R}_0>1$ and die out otherwise.
	
\begin{figure}
		\epsfig{file=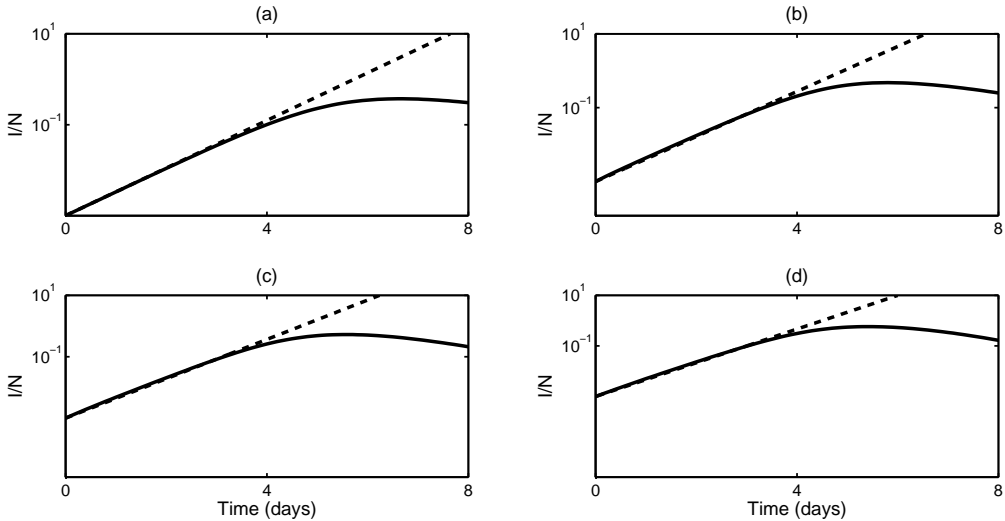,width=16cm}
		\caption{Proportion of infected individuals during a boarding school influenza outbreak with $\beta=1.66$/day and $\gamma=0.4545$/day (Keeling and Rojani 2008) \cite{keeling2008modeling}. In each plot the solid black line is the numerical solution of the model (\ref{eq:stage_ode}) with an appropriate number of stages, and the dashed line is the exponential growth curve with the rate determined by equation (\ref{eq:R0lambda}) shown on a logarithmic scale. (a) One-stage model with $\lambda\approx 1.2055$, (b) Two-stage model with $\lambda\approx 1.4035$, (c) Three-stage model with $\lambda\approx 1.4762$, (d) Five-stage model with $\lambda\approx 1.534$. In each case note that in the earliest stages the exponential approximation is virtually identical to the infection curve.}
		\label{fig:approxcurves}
	\end{figure}

	The basic reproduction number for the system (\ref{eq:stage_ode}) can be found as follows (Hyman and Stanley 1999; Ma and Earn 2006; van den Driessche and Watmough 2002) \cite{hyman99,ma2006generality,van2002reproduction}
	\begin{equation}
	\mathcal{R}_0 = \frac{\beta}{\gamma},
	\end{equation}
	which depends on the average duration of infection $1/\gamma$ but is independent of the number of stages in the model. A practically important characteristic of an epidemic outbreak is the {\it final epidemic size} (Keeling and Rojani 2008) \cite{keeling2008modeling}. Since the total population size is closed with no inflow or outflow of individuals, i.e. $S(t)+I_1(t)+I_2(t)+...+I_K(t)+R(t)=1$, at the end of an outbreak we have a burn-out of the epidemic, i.e. $I_1=I_2=\ldots=I_K=0$, and hence $S(\infty)+R(\infty)=1$ and $R(\infty) = 1 - S(\infty)$. This results in the following implicit equation for the final size of an epidemic that determines the proportion of individuals not affected by the disease (Anderson and May 1992; Diekmann and Heesterbeek 2000) \cite{AM92,DieHee}
	\begin{equation}
	R(\infty) = 1 - e^{-\mathcal{R}_0 R(\infty)},
	\label{eq:finalsize}
	\end{equation}
	which coincides with the final epidemic size in the original $SIR$ model (Kermack and McKendrick 1927) \cite{SIR}. Ma and Earn (2006) \cite{ma2006generality} have recently discussed various aspects related to the derivation and validity of formula (\ref{eq:finalsize}), and Andreasen (2011) \cite{Andr11} has studied the effects of population heterogeneity on the size of epidemic. A major implication of the above results is the fact that inclusion of possibly more realistic gamma distribution of infectious periods does not alter the threshold of an epidemic outbreak, nor does it affect the final epidemic size. One should note, however, that when a stochastic version of the $SI^KR$ model is considered, the number of stages influences the distribution of final epidemic sizes, while the average final size remains the same (Black and Ross 2015; House et al. 2013) \cite{black,house13}.
	\begin{figure}
		\epsfig{file=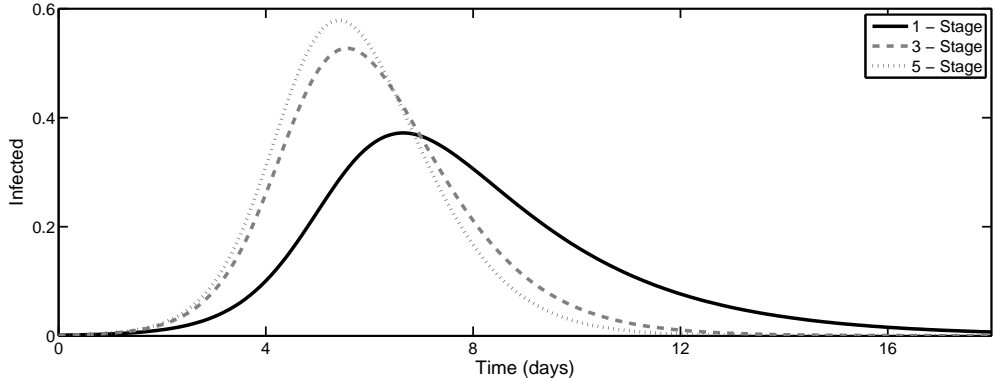,width=16cm}
		\caption{A comparison of infection dynamics for a one-, three- and five-stage $SI^KR$ models with data from (Keeling and Rojani 2008) \cite{keeling2008modeling}. Each curve represents the sum of all $I_i$ in the model. Adding extra stages causes epidemics to occur earlier and result in a higher peak of epidemic, although $\mathcal{R}_0$ and the final size are identical for each curve. The parameter values are $N = 1000$, $\beta = 1.66/$day, $\gamma = 0.4545/$day.}
		\label{fig:135compare}
	\end{figure}
	We see that in Fig.~\ref{fig:135compare} the three curves show that considering multi-stage infectious periods has a significant effect of the dynamics of the epidemic. In order to get a better understanding of the distinction in the dynamics of $SIR$ and $SI^KR$ models, it is therefore instructive to look at the development of epidemics. In the standard $SIR$ model, an outbreak can only take place if $\mathcal{R}_0 >1$, and at the initial stage, the number of infected individuals can be approximated as $I(t)\approx I(0)\exp(\lambda t)$, where the growth rate is $\lambda=\gamma(\mathcal{R}_0 -1)$. In the case of a multi-stage $SI^KR$ model, however, the basic reproduction number $\mathcal{R}_0$ does not depend on the number of stages, hence, it cannot by itself be used to determine the exponential growth rate during an early stage of an outbreak. For this model Wearing et al. (2005) \cite{wearing2005appropriate} have derived the following relation between the basic reproduction number $\mathcal{R}_0$ and the initial growth rate $\lambda$
	\begin{equation}
	\mathcal{R}_0 = \frac{\lambda}{\gamma\left(1-\left(\frac{\lambda}{K\gamma}+1\right)^{-K}\right)}.
	\label{eq:R0lambda}
	\end{equation}
	Figure~\ref{fig:approxcurves} illustrates early dynamics of epidemic outbreaks for different numbers of stages; in each case an exponential curve was fitted, which provides an accurate approximation for the initial growth rate of the infection as determined by equation (\ref{eq:R0lambda}). This figure shows the effects of the gamma distribution on the early growth rate, peak prevalence and overall time frame of the disease, and it also suggests that the largest effect of the gamma distribution on the disease dynamics occurs during intermediate stages of disease progression.
	
	Besides the basic reproduction number, final epidemic size and the initial growth rate of an epidemic, another practically important characteristic of epidemic outbreaks is the peak prevalence defined as the maximum number or proportion of infected individuals that can be achieved during an outbreak. In the case of an $SIR$ model, the peak prevalence can be found as follows (Feng 2007; House and Keeling 2011b) \cite{feng2007final,HK11}
	\[
	\displaystyle{I_{max}=1-\frac{1}{\mathcal{R}_0}[1+\ln(\mathcal{R}_0)].}
	\]
	Feng (2007) \cite{feng2007final} has recently considered an SEIR model with gamma distributed infectious period and derived an expression for the peak of a weighted average of infectious compartments. This result gives some intuition into how the number of stages affects peak prevalence, but it does not provide a closed form expression for the actual peak prevalence in an $SI^KR$ model. Numerical results in Fig.~\ref{fig:approxcurves} suggest that for the same average infectious period, the overall peak prevalence increases with the number of stages included in the model.
	
	\section{Network dynamics with multiple stages}
	
	Inclusion of multiple stages of infection in the $SI^KR$ model gives a more realistic representation of the infectious period, but the model still has certain limitations due to its underlying assumptions. In the model (\ref{eq:stage_ode}) it is assumed that the disease is not fatal, and that transitions between different infected classes, or stages of infection, take place at exactly the same rate $K\gamma$. Another major assumption behind model (\ref{eq:stage_ode}) is that the population is well-mixed, i.e. each individual has equal chances of encountering and transmitting a disease to any other individual in a population. Whilst this may be appropriate in the case of outbreaks in small closed communities, for a large number of communicable diseases, such as SARS, influenza and most sexually transmitted infections, this assumption is a gross simplification of the actual dynamics as it overlooks spatial variability, as well as the complexities of a network structure for infections that are transmitted through direct close contact between individuals (House and Keeling 2011a; Keeling and Eames 2005) \cite{house2011insights,keeling_eames}.
	
	Modelling complex contact patterns explicitly via networks has had a profound effect on mathematical epidemiology. This new modelling framework has led to a myriad of models ranging from exact to mean-field and simulation models (Pastor-Satorras et al. 2014; Danon et al. 2011; Keeling and Eames 2005; Newman 2003; Boccaletti et al. 2006)\cite{pastorreview, danon, keeling_eames, newmanreview, boccaletti2006}. The many degrees of freedom in modelling offered by networks however, often comes at the price of increasing levels of complexity, where models can be challenging to evaluate analytically and sometimes even numerically. Nevertheless, many valuable paradigm models have been developed which have furthered our understanding of the impact of contact heterogeneity, preferential mixing and clustering on the outbreak threshold and other epidemic descriptors. A particularly useful way of capturing epidemic dynamics on networks is by using the pairwise model (Keeling 1999) \cite{keeling1999}. This model is based around deriving in a hierarchical way evolution equations for the expected number of nodes, edges, triples and so on. A closure is then employed that curtails the dependence on ever higher order moments. Its premise is simple and quite intuitive, although it can be also shown rigorously (Taylor et al. 2012) \cite{Markpairwise} that pairwise models before closure are exact. The basic idea of the model is to recognise that changes at node level depend on the status of the neighbours and thus involves edges, e.g. the rate of change in the number of infectious nodes is proportional to the number of $S-I$ links in the network. Similarly, the number of edges can change due to pair interactions and transitions but also due to interactions induced from outside the edge, e.g. the number of $S-S$ links decrease proportionally to the number of $S-S-I$ triples, where infection from the $I$ node destroys the fully susceptible pair. This framework has been used and extended extensively, to asymmetric (Sharkey et al. 2006) \cite{sharkey06} and weighted networks (Rattana et al. 2013) \cite{rattana2013class} for example, and has proved to be a valuable framework.
	
	\subsection{Pairwise model}
	
	As a first step in the analysis of dynamics of multi-stage epidemics on networks, we re-formulate the $SI^KR$ model using the framework of \emph{pairwise equations}, which allows one to analyse the expected values for the number of nodes and links of each type as a function of time (Keeling 1999; House and Keeling 2011a; Taylor et al. 2012) \cite{keeling1999,house2011insights,Markpairwise}. The particular strength of pairwise models lies in their analytical tractability and the fact that they provide a more accurate description than well-mixed ODE models but do not go to the level of full individual-based stochastic simulations (House and Keeling 2011a) \cite{house2011insights}. In this formalism of pairwise models, notations $[X]$, $[XY]$ and $[XYZ]$ are used to denote the expected numbers of individuals in state $X$, the expected number of links between nodes of type $X$ and $Y$ and the expected number of triples of the form $X-Y-Z$, respectively. More precisely, given a `frozen' network with nodes labels $X$, $Y$ or $Z$ and subscripts indicating nodes $i, j$ and $k$ then 
	\[
		[X]=\sum_{i=1}^{N}X_i, \quad [XY]=\sum_{i,j=1}^{N}X_iY_jg_{ij}, \quad [XYZ]=\sum_{i,j,k=1}^{N}X_iY_jZ_kg_{ij}g_{jk},
	\]
	where $X, Y, Z \in \{S, I_1, I_2, \ldots, I_K, R\}$, and $G=(g_{ij})_{i,j=1,2,\dots,N}$ is the adjacency matrix of the network such that $g_{ii}=0$, $g_{ij}=g_{ji}$ and $g_{ij}=g_{ji}=1$ if nodes $i$ and $j$ are connected and zero otherwise. Moreover, $X_i$ returns one if node $i$ is in state $X$ and zero otherwise. The average degree of each node is denoted by $n$, and the number of nodes in the network by $N$.  The new pairwise $SI^KR$ model with a gamma distributed infectious period can then be written as follows,
	\begin{equation}\label{eq:ex_pair}
	\begin{array}{l}
	\dot{[S]}   = -\tau \sum_{i=1}^K{[SI_i]}, \\\\
	\dot{[I_1]} = \tau \sum_{i=1}^K{[SI_i]} - K\gamma [I_1], \\\\
	\dot{[I_j]} = K\gamma [I_{j-1}] - K\gamma [I_j], \quad \mbox{for} \quad j=2, 3, \ldots, K, \\\\
	\dot{[SS]}= -2\tau\sum_{i=1}^K[SSI_i], \\\\
	\dot{[SI_1]}= -(\tau+K\gamma)[SI_1] + \tau\left(\sum_{i=1}^K{[SSI_i]} - \sum_{i=1}^K{[I_iSI_1]}\right), \\\\
	\dot{[SI_j]}= -(\tau+K\gamma)[SI_j] + K\gamma[SI_{j-1}] -\tau \sum_{i=1}^K{[I_iSI_j]}, \quad \mbox{for} \quad j=2, 3, \ldots, K.
	\end{array}
	\end{equation}
	where $\tau=\beta/n$ is the transmission rate per link. Since we consider a closed population, this immediately implies $[S] + \sum_{i=1}{K}[I_i] + [R] =N$. The system (\ref{eq:ex_pair}) is not closed as additional equations describing the dynamics of triples are needed. To eliminate this dependence on higher moments and close the system, we will use the classical moment closure approximation which assumes that short loops and clusters are excluded from the network and that there is no correlation between nodes with a common neighbour (Keeling 1999) \cite{keeling1999}.
	\begin{equation}\label{eq:closure_pair}
	\begin{array}{l}
	\displaystyle{[SSI_i]   \approx \frac{(n-1)}{n}\frac{[SS][SI_i]}{[S]}, \quad \mbox{for} \quad i=1:K,}\\ \\
	\displaystyle{[I_jSI_i] \approx \frac{(n-1)}{n}\frac{[I_jS][SI_i]}{[S]}, \quad \mbox{for} \quad i,j=1:K.}
	\end{array}
	\end{equation}
	Applying these closures to the system (\ref{eq:ex_pair}) makes it a self-consistent system of $(2K+2)$ equations.
	
	\subsection{The probability of transmission across an infected edge}
	\label{sec:tautilde}
	
	When one considers a stochastic network-based simulation, an important quantity characterising the disease dynamics is the probability $\tilde{\tau}$ of disease transmission across a given $S-I$ link. In a simple one-stage model, where both infection and recovery are assumed to be distributed exponentially, the probability of no infection event occurring during time $t$ is given by $p_0(t)=e^{-\tau t}$; hence $1 - p_0(t)$ is the probability that infection does take place over the same time period. Averaging this via integration for all possible recovery times yields the probability that the susceptible node becomes infected. In a standard $SIR$ model with exponentially distributed infectious and recovery period, this probability is, therefore (Danon et al. 2011; Diekmann, De Jong and Metz 1998) \cite{danon, diekmann98}
	\begin{equation}
	\tilde{\tau} = 1 - \frac{\gamma}{\tau + \gamma} = \frac{\tau}{\tau+\gamma}.
	\label{eq:tilde}
	\end{equation}
	In the case of an $SI^KR$ model, the duration of infection is described by the density function of the appropriate gamma distribution
	\begin{equation}
	g(x;K,1/(K\gamma)) = \frac{1}{(K-1)!}(K\gamma)^K x^{K-1} e^{-K\gamma x}.
	\label{eq:gamma_tau}
	\end{equation}
	The implication of this fact is the following result for the probability of transmission across an edge.\\
	\noindent {\bf Lemma1.} {\it For the stochastic $SI^KR$ model with the period of infection following the gamma distribution (\ref{eq:gamma_tau}), the probability of disease transmission across a given $S-I$ link is given by
		\begin{equation}
		\tilde{\tau} = 1 - \left(\frac{K\gamma}{\tau + K\gamma}\right)^K.
		\label{eq:Kstage}
		\end{equation}
	}
	\noindent The proof of this lemma is given in Appendix A.\\
	
\begin{figure}
		\hspace{2cm}
		\epsfig{file=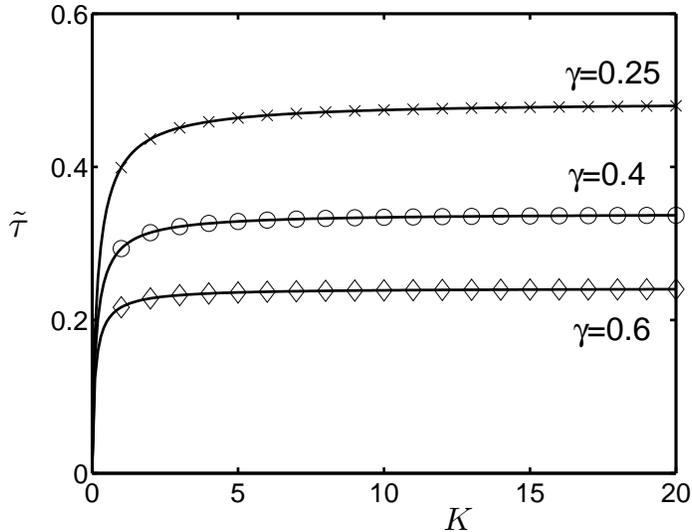,width=10cm}
		\caption{Dependence of the probability of transmission across an $S-I$ edge $\tilde{\tau}$ on the number of stages $K$ as given by Eq. (\ref{eq:Kstage}) for different mean infectious periods with $\tau=0.166$. Crosses, circles and diamonds correspond to integer values of $K$ on each curve.}
		\label{fig:taulimit}
	\end{figure}

	By rewriting expression (\ref{eq:Kstage}) in the form
	\[
	\tilde{\tau} =1 - \left(\frac{K\gamma + \tau - \tau}{\tau + K\gamma}\right)^K = 1 - \left(1 - \frac{\tau}{\tau + K\gamma}\right)^K,
	\]
	and using the fact that $e^x = \lim_{n \to \infty} \left(1+x/n\right)^n$, it follows that
	\begin{equation}
	\displaystyle{\lim_{K \to \infty} \tilde{\tau}(K) = 1 - \exp\left(-\frac{\tau}{\gamma}\right)}.
	\label{eq:taulimit}
	\end{equation}
	Figure~\ref{fig:taulimit} illustrates the dependence of $\tilde{\tau}$ on the number of stages $K$, as well as a limiting behaviour as $K\to\infty$.
	This figure illustrates that while $\tilde{\tau}$ is growing with the increasing number of stages $K$, it eventually saturates at a level determined by Eq. (\ref{eq:taulimit}). In fact, this saturation at higher $K$ is observed not only in the probability of transmission, but also in the peak prevalence rate, as well as in the early growth rate. When compared to an exponential distribution, it is these substantial changes in $\tilde{\tau}$ observed for smaller values of $K$ that explain the changes in the profile of the infection curves. As will be shown later, $\tilde{\tau}$ is a very important quantity that controls various properties of epidemic dynamics, such as the threshold for an outbreak and the final size of an epidemic.
	
	\subsection{$\mathcal{R}_0$-like threshold parameter}
	
Unlike epidemic models in well-mixed populations, defining an appropriate $\mathcal{R}_0$ for
pairwise models is more challenging. This is in part due to the difficulty of identifying the
 ÔtypicalÕ infectious individual. In order to derive a value for $\mathcal{R}_0$, one needs to
consider and correctly account for the correlation between susceptible and infected
nodes and measure $\mathcal{R}_0$ when this has stabilised, see Keeling (1999) \cite{keeling1999} and Eames
(2008) \cite{eames08}. Intuitively, this means that the epidemic is allowed to spread in order
to become established in the network. This allows for Ôtypical' infectious individuals
to develop and for $\mathcal{R}_0$ to be measured. In large networks this regime can still be
considered to be close to or only a small perturbation away from the disease-free
steady state.
	
	We now proceed to derive an $\mathcal{R}_0$-like threshold parameter $\mathcal{R}$ which can be used to predict when the epidemics occur, by allowing outbreaks only when $\mathcal{R} > 1$ (Rattana et al. 2013) \cite{rattana2013class}. To this end, we linearise the model (\ref{eq:ex_pair}) with a classic closure (\ref{eq:closure_pair}) near the disease-free equilibrium which has the form $[S]=N$, $[SS]=nN$, and all other quantities being zero. As in the standard approach, the condition necessary for the initial growth of an epidemic is that the dominant eigenvalue $\lambda_{max}$ of the resulting characteristic polynomial is real and positive, and a threshold parameter is obtained as a condition on system parameters that ensure the stability change, i.e. $\lambda_{max} = 0$. In the Appendix B it is shown that the characteristic equation for eigenvalues $\lambda$ of the linearised system near the disease-free steady state for a $K$-stage model (\ref{eq:ex_pair}) is given by
	\[
	\begin{array}{l}
		\lambda^2(\lambda + K\gamma)^K\Bigg[(\tau + K\gamma + \lambda)^K - \tau(n-1)\Bigg[(\tau + K\gamma + \lambda)^{K-1}\Bigg.\Bigg. \\
		\hspace{2cm} \left.\left.+ \sum_{i=1}^{K-1}(K\gamma)^{K-i}(\tau + K\gamma + \lambda)^{i-1}\right]\right]=0.
	\end{array}
	\]
	In Appendix B we prove that the largest eigenvalue $\lambda$ satisfying this equation goes through zero, i.e. $\lambda_{max} = 0$, when
	\begin{equation}
	\mathcal{R} := (n-1)\tilde{\tau} = 1.
	\label{eq:R}
	\end{equation}
	This defines a new $\mathcal{R}_0$-like threshold parameter with $\tilde{\tau}$ introduced in (\ref{eq:Kstage}). A closer inspection shows that this parameter $\mathcal{R}$ describes the probability of spreading the disease across a given link multiplied by the likely number of susceptible contacts of the individual assuming that they are the earliest people being infected, which perfectly agrees with the standard definition of $\mathcal{R}_0$ as the average number of secondary cases produced in a fully susceptible population by a single typical infectious individual. Whilst $\mathcal{R}$ does not quantify the early growth rate of an epidemic, through its dependence on $\tilde{\tau}$ and $K$ it allows one to better predict epidemic outbreaks in the case of a more realistic gamma distribution of infectious period, where in the case of an exponential distribution with the same mean infectious period. We also note that whilst in the implementation of the classic $SI^KR$ model there was no effect of changing the number of stages on $\mathcal{R}_0$, this more sophisticated model results in a threshold which implicitly accounts for multi-stage infectious periods.
	
	\subsection{The final size of an epidemic}
	\label{sec:finalsize}
	
	Since the pairwise model (\ref{eq:ex_pair}) is a network representation of an epidemic with life-long immunity and fixed population size, eventually an epidemic will burn out, leaving some proportion of the population unaffected and still susceptible to the disease. Since $[I_1](\infty)=[I_2](\infty)=\ldots=[I_K](\infty)=0$, the final size of an epidemic is given by the proportion of people in the removed class, i.e. $[R]_{\infty} = N-[S]_{\infty}$. As we saw earlier for the $SI^KR$ model (\ref{eq:stage_ode}) in a well-mixed population, the final size of a single epidemic does not change with the number of stages. However, the same conclusion no longer holds for the pairwise model (\ref{eq:ex_pair}) with the closure (\ref{eq:closure_pair}), in which case we have the following result.
	
	\begin{thm}
		\label{thm:final_size}
		For a single epidemic outbreak in a closed population with a vanishingly small starting level of infection, the final size of an epidemic in the pairwise model (\ref{eq:ex_pair}) with the classical closure (\ref{eq:closure_pair}) is given by
		\begin{equation}
		R_{\infty} = 1 - \left( 1 - \tilde{\tau} + \tilde{\tau}\theta\right)^n,
		\label{eq:Keelingsize}
		\end{equation}
		\noindent where
		\begin{equation}
		\theta = \left( 1 - \tilde{\tau} + \theta\tilde{\tau}\right)^{n-1},
		\label{eq:Keelingsizetheta}
		\end{equation}
		and $\tilde{\tau}$ is defined in (\ref{eq:Kstage}).
	\end{thm}
	
	\noindent {\bf Proof.} To prove this statement we extend the methodology developed by Keeling (1999) \cite{keeling1999} for one-stage epidemics. We first introduce some new variables and parameters
	\[
	a = \frac{n-1}{n}, \quad F = \frac{\sum_{i=1}^K [SI_i]}{[S]^a}, \quad G = \frac{[SR]}{[S]^a}, \quad L = \frac{[SS]}{\exp(n[S]^{1/n})[S]^{2a}}, \quad M = \frac{[SS]}{[S]^a},
	\]
	and
	\begin{equation}
	P_i = \frac{[SI_i]}{[S]^a} \quad \mbox{for } \, i=1,2, \ldots, K.
	\label{eq:P's}
	\end{equation}
	From (\ref{eq:ex_pair}) and the easily derived function
	\[
	\dot{[SR]} = -\tau \frac{[SR]\sum_{i=1}^K [SI_i]}{[S]} + K\gamma [SI_K],
	\]
	it follows that these new variables satisfy the following system of equations
	\begin{equation}\label{eq:Fdot}
	\begin{array}{l}
	\displaystyle{\dot{F}=-\tau F - K\gamma P_K + a\tau\frac{[SS]}{[S]}F,}\\\\
	\displaystyle{\dot{G}= K\gamma P_K,}\\\\
	\displaystyle{\dot{L}= -a\tau\frac{[SS]}{[S]}F,}\\\\
	\displaystyle{\dot{M}= \tau MF.}
	\end{array}
	\end{equation}
	Since $[I_i](0)=[I_i](\infty)=0$ for any $i=1,É,K$, this implies $F(0)=F(\infty)=0$. Integrating the first equation in (\ref{eq:Fdot}) gives
	\begin{equation}\label{eq:F}
	\begin{array}{l}
	\displaystyle{F(\infty) - F(0)= 0= -\tau \int_0^{\infty}F dt - K\gamma \int_0^{\infty} P_K dt + a\tau\int_0^{\infty}\frac{[SS]}{[S]}F dt}\\\\
	= -[\ln(M(\infty)) - \ln(M(0))] - [G(\infty) - G(0)] - \left[L(\infty)-L(0)\right]\\ \\
	= -[\ln(M(\infty)) - \ln(M(0))] -  \tilde{\tau}\left[L(\infty)-L(0)\right],
	\end{array}
	\end{equation}
	where in the last step we have used the fact that $G(0)=0$ and the relation 
	\begin{equation}\label{eq:SR}
	G(\infty) = \frac{[SR]_{\infty}}{[S]_{\infty}^a} = (\tilde{\tau}-1)[L(\infty) - L(0)],
	\end{equation}
	\begin{figure}
		\epsfig{file=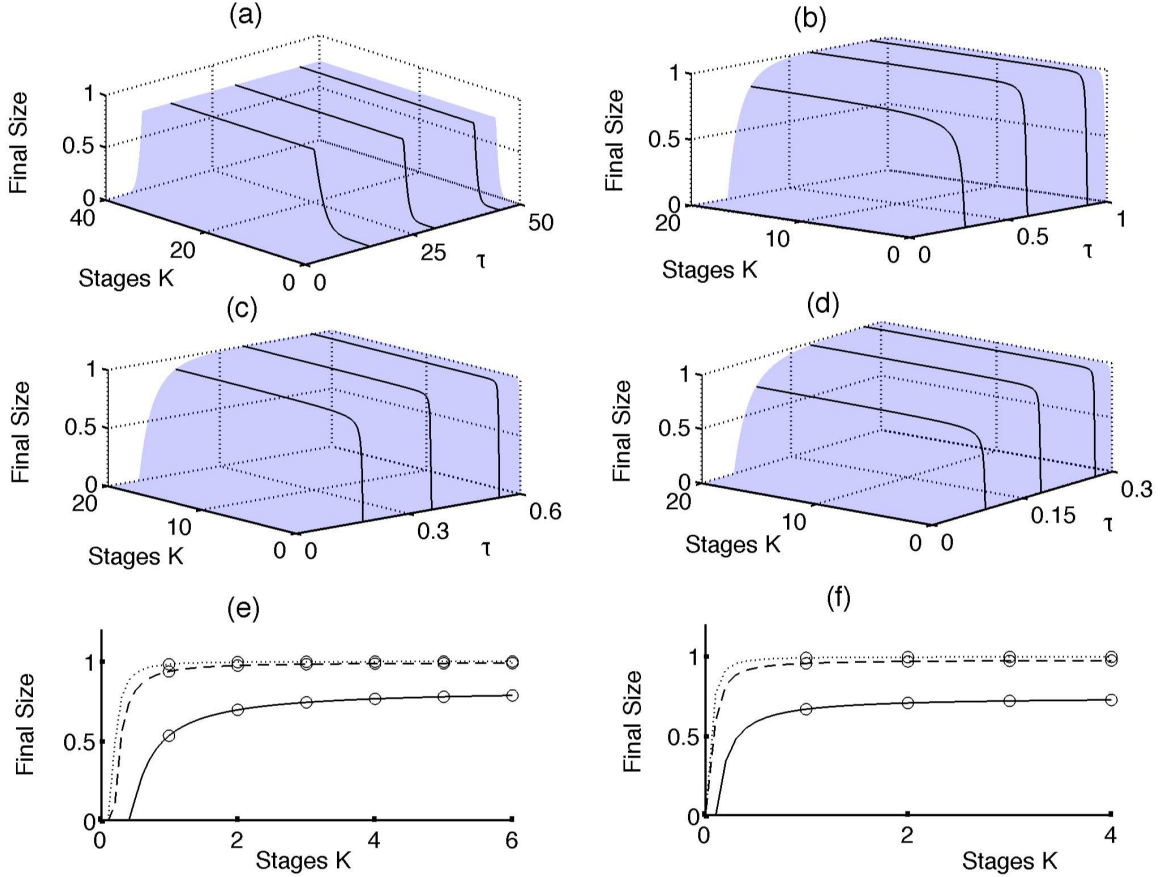,width=16cm}
		\caption{Dependence of the final size of an epidemic (\ref{eq:Keelingsize}) on the per-link transmission rate $\tau$ and the number of stages $K$ in the pairwise model (\ref{eq:ex_pair}) with $\gamma=0.4545$ for different average node degrees. (a) $n=2$. (b) $n=4$. (c) $n=7$. (d) $n=10$. (e) $n=4$, $\tau=0.3$ (solid), $\tau=0.6$ (dashed), $\tau=0.9$ (dotted). (f) $n=10$, $\tau=0.09$ (solid), $\tau=0.18$ (dashed), $\tau=0.27$ (dotted). Circles correspond to integer values of $K$ on each curve. The case $n=2$ is used solely for illustrative purposes, as the resulting networks would be disconnected and thus inappropriate for direct comparison to results from the pairwise model.}
		\label{fig:finalsizes}
	\end{figure}
	derived in Appendix C together with another relation 
	\begin{equation}\label{eq:SS}
	[SS]_{\infty} = \frac{n[S]_{\infty}^{2a}}{N^{a-1/n}}.
	\end{equation}
	Substituting these two relations into Eq.~(\ref{eq:F}) and using the fact that $[S](0)=N$ yields
	\[
	\begin{array}{l}
	\displaystyle{0= nN^{1/n} -n[S]_{\infty}^{1/n} - \tilde{\tau}\left(\frac{n[S]_{\infty}^a}{N^{a-1/n}}-nN^{1/n}\right)}\\\\
	\displaystyle{= 1 - \left(\frac{[S]_{\infty}}{N}\right)^{1/n} - \tilde{\tau}\left[\left(\frac{[S]_{\infty}}{N}\right)^a-1\right].}
	\end{array}
	\]
	Introducing the fraction of susceptible individuals as $S_{\infty}=[S]_{\infty}/N$, the above equation can be rewritten as follows,
	\[
	\displaystyle{1 - S_{\infty}^{1/n}= \tilde{\tau}\left(S_{\infty}^a -1\right),}				
	\]
	or alternatively, as another implicit equation for $S_{\infty}$
	\begin{equation}\label{Sinf}
	\displaystyle{S_{\infty}= \left(1 - \tilde{\tau} + \tilde{\tau}\theta\right)^n,\hspace{0.5cm}\mbox{where }\theta = S^a_{\infty}}.
	\end{equation}
	Since $[I]_i(\infty)=0$, introducing $R_{\infty}=[R]_{\infty}/N$ yields the desired expression for the final size of an epidemic
	\[
	\displaystyle{R_{\infty}=1-S_{\infty}=1-\left(1 - \tilde{\tau} + \tilde{\tau}\theta\right)^n.}
	\]
	Using the fact that $\theta = S^a_{\infty}$, equation (\ref{Sinf}) can be rewritten in the form
	\[
	\displaystyle{\theta^{1/a}=\left(1 - \tilde{\tau} + \tilde{\tau}\theta\right)^n\hspace{0.3cm}\Longrightarrow \theta=\left(1 - \tilde{\tau} + \tilde{\tau}\theta\right)^{n-1},}
	\]
	where in the last step we have used the relation $a=(n-1)/n$. This completes the proof.
\hfill$\blacksquare$

	\bigskip
	We note that our result in Theorem~\ref{thm:final_size} is functionally identical to the result achieved by Keeling (1999) \cite{keeling1999}, and it generalises the final size equation by replacing $\tau/(\tau + \gamma)$ with the parameter $\tilde{\tau}$. In the case $K=1$ these two values are equivalent, thus we have perfect agreement with the existing theory. Equivalent relations have also been derived by Newman (2002) \cite{newmannetworks} using percolation theory. Those results were later corrected and shown to hold in all cases where the distribution of infectious periods is degenerative (Kenah and Robins 2007) \cite{kenahrobins}. An equivalent relation has been derived for a static configuration network model with an arbitrary degree distribution (Miller 2012) \cite{miller}. Figure~\ref{fig:finalsizes} illustrates Theorem~\ref{thm:final_size} by showing how the final size of an epidemic on a network depends on the number of infectious stages and, hence, the shape of the distribution of infectious period, which makes it different from earlier analytical results for a well-mixed population (Ma and Earn 2006) \cite{ma2006generality}. This suggests that inclusion of a more realistic population structure has effect not only on the intermediate disease dynamics, but also on the final proportion of the population that will be affected by the disease. Furthermore, this Figure suggests that for the same mean infectious period, the final size of an epidemic is increasing with the increasing number of stages $K$. One should note that the number of stages $K$ has the largest effect on the final size of an epidemic for sufficiently low values of $K$, and then this dependence saturates. As expected, the average node degree $n$ plays an important role, with the minimum value of $\tau$ or $K$ required for an epidemic outbreak decreasing with increasing $n$ in perfect agreement with an earlier result in Eq. (\ref{eq:R}). Stochastic simulations (not shown) demonstrate excellent agreement with the results in Fig.~\ref{fig:finalsizes}, especially for denser networks. The conclusions of Theorem~\ref{thm:final_size} highlight the importance of collecting accurate and reliable data about the infectivity profile of a disease for predicting the scale of an outbreak.
	
	\begin{figure}
		\hspace{1cm}
		\epsfig{file=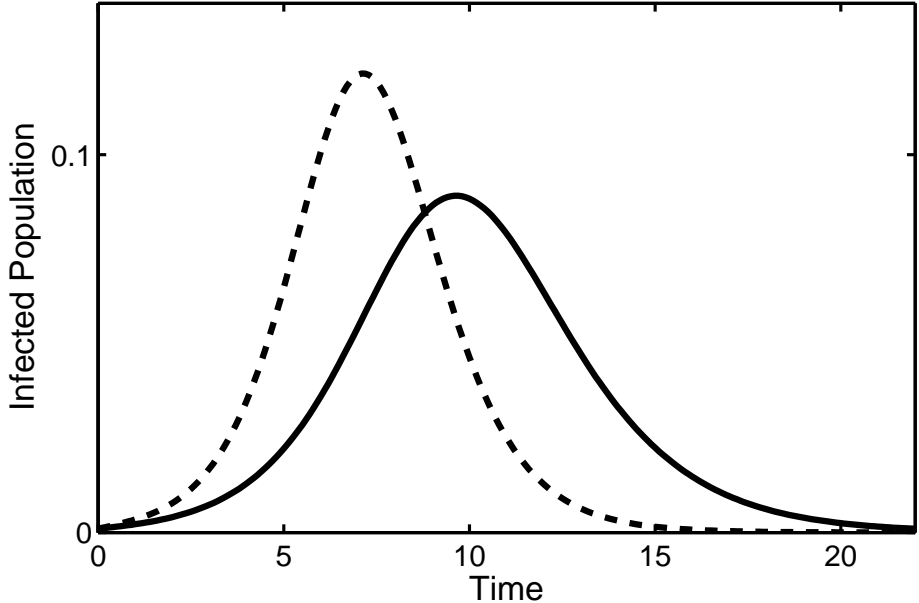,width=14cm}
		\caption{Numerical solution of the pairwise $SI^KR$ model (\ref{eq:ex_pair}) with different average infectious periods and a different number of stages, but the same final size due to identical transmissibility $\tilde{\tau}$. Parameter values are $\tau = 0.2$,  $n=10$, $\gamma = 1$ and $K=1$ (solid) and $\tau = 0.2$,  $n=10$, $\gamma \approx 1.06$ and $K=3$ (dashed). The solution curves for the overall infected population show a radically different intermediate behaviour, but with $\tilde{\tau} = 1/6$ in both cases, they have the same final epidemic size.}
		\label{fig:fs_same}
	\end{figure}
	
	It is worth noting that whilst the final size depends on the distribution of the infectious period, this dependence is not necessarily unique. This means that two different distributions of infected periods can provide the same transmissibility $\tilde{\tau}$, resulting in the same final epidemic size in accordance with Theorem~\ref{thm:final_size} but having different intermediate dynamics of infection, as illustrated in Fig.~\ref{fig:fs_same}. The consequence of this observation is that although the epidemic threshold and final epidemic size can both be accurately computed using an estimate for the transmissibility of the disease (Newman 2002) \cite{newmannetworks}, it is not sufficient to correctly predict the dynamics of the infection spreading process over time, which can be done with our model.
	
	\section{Impact of a realistic infectious period distribution: case studies}
	
	In order to test the accuracy of the pairwise model (\ref{eq:ex_pair}) and to illustrate the role played by the distribution of infectious period, we consider the examples of outbreaks of several diseases mentioned in Table \ref{tab:distributions} in a population that is initially fully susceptible. We concentrate on two common and fairly simple network structures, namely, homogeneous and Erd{\H o}s-R{\'e}nyi networks (Newman 2010) \cite{networks}, with stochastic simulations being performed using a Gillespie algorithm (Gillespie 1977; Chen and Bokka 2005) \cite{Gil77,CB05}. We restrict our attention to these network types as we have a homogeneous pairwise model and we would not expect it to work well for other networks. Following the derivation of the pairwise model, the per-link transmission rate is taken to be $\tau=\beta/n$, and we now perform the comparison of an average of 250 stochastic outcomes of serious epidemics on a homogeneous and Erd{\H o}s-R{\'e}nyi networks against the results of a pairwise model with gamma distributed infectious period. To highlight the impact of including a realistic distribution for the infectious period, we compare the results of simulations with realistic values of parameters from Table~\ref{tab:distributions} against those obtained using an exponentially distributed infectious period as assumed in many existing models.

\begin{figure}
		\hspace{-1cm}
			\epsfig{file=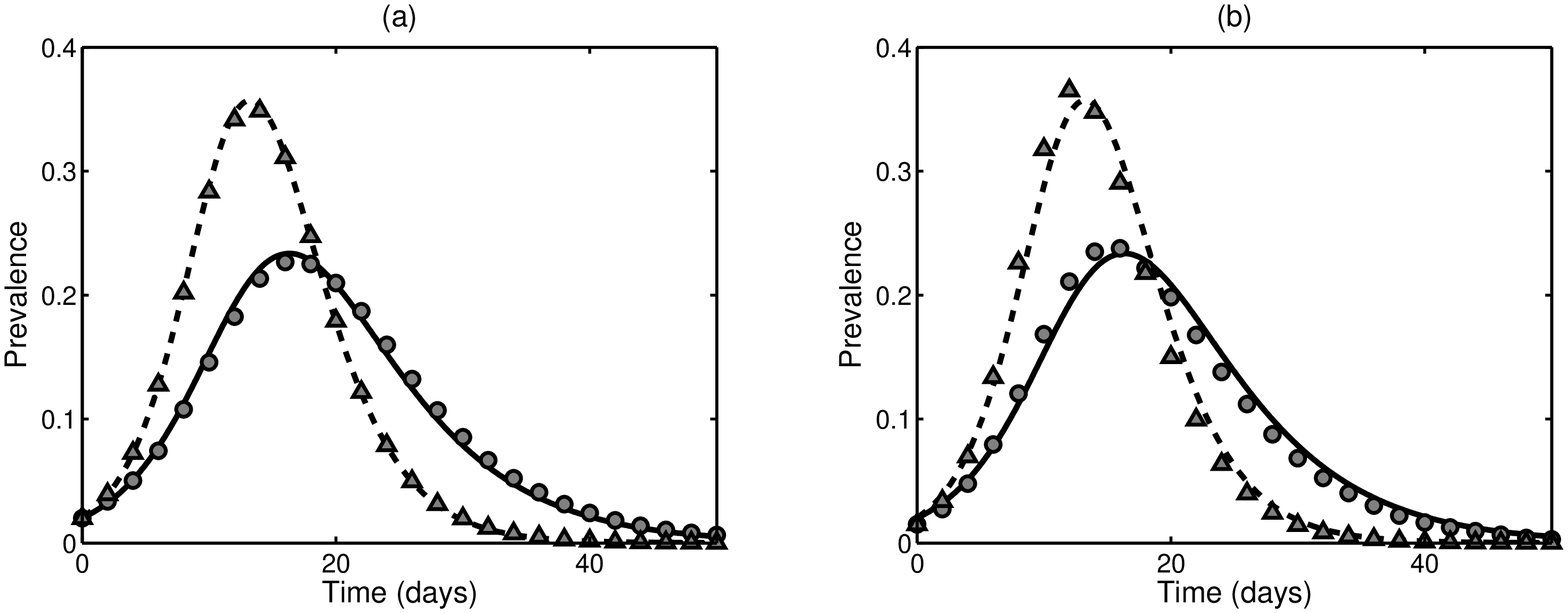,width=18cm}
			\caption{Simulation of a SARS outbreak using data from Table~\ref{tab:distributions} with $n=10$ and $N=1000$. Lines correspond to a numerical solution of the pairwise model (\ref{eq:ex_pair}) ($K=1$ solid line, $K=3$ dashed line), while symbols represent the average of 250 serious outbreaks ($K=1$ filled circles, $K=3$ triangles). (a) Homogeneous network. (b) Erd{\H o}s-R{\'e}nyi random graph.}
			\label{fig:SARS}
		\end{figure}
	
	Severe Acute Respiratory Syndrome (SARS) is a viral disease characterised by flu-like symptoms which is primarily spread through close contacts with infected individuals that makes it a perfect candidate for deducing some basic parameters from epidemiological observations. Figure~\ref{fig:SARS} illustrates the comparison of SARS dynamics on homogeneous and Erd{\H o}s-R{\'e}nyi networks with a pairwise approximation. One can observe that the effects of including more stages in the disease model on intermediate behaviour are similar to those seen earlier, namely, that gamma distribution of infectious period shortens the overall duration an epidemic and increases peak prevalence. It is also worth noting that, in accordance with Theorem~\ref{thm:final_size}, the final size of an epidemic also increases with $K$.
	
\begin{figure}
		\hspace{-1.cm}
		\epsfig{file=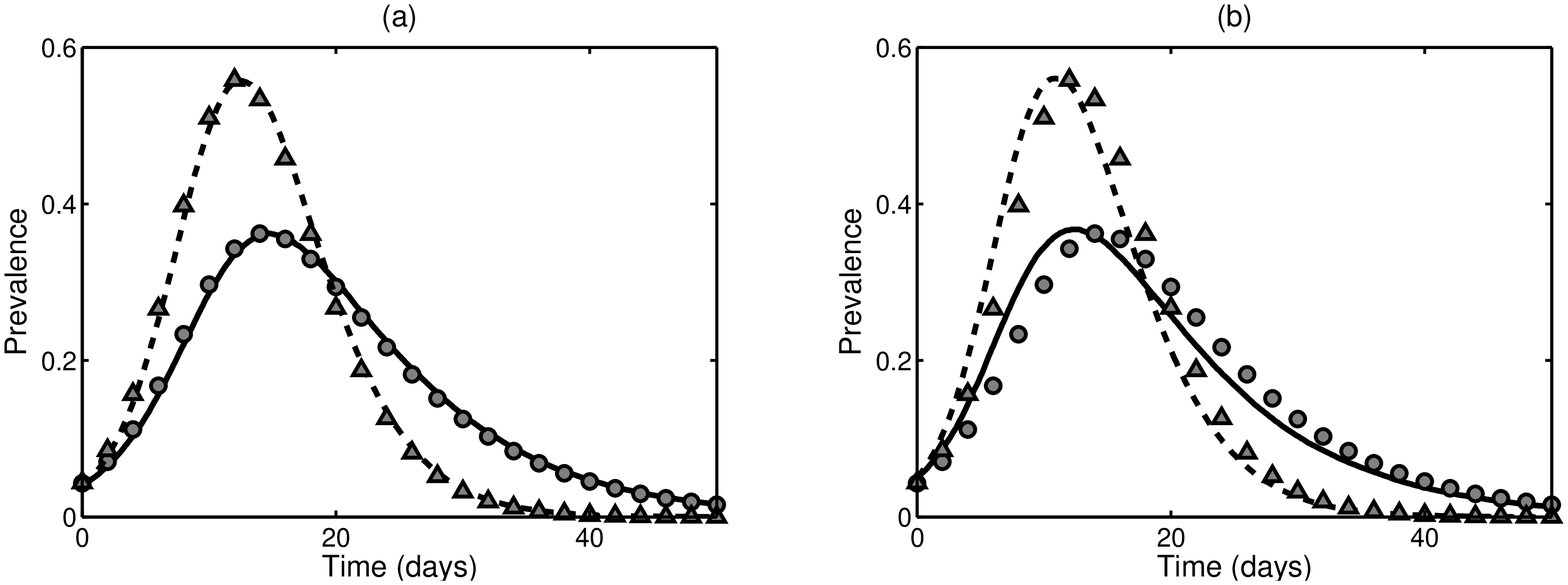,width=18cm}
		\caption{Simulation of a smallpox outbreak using data from Table~\ref{tab:distributions} with $n=10$ and $N=1000$. Lines correspond to a numerical solution of the pairwise model (\ref{eq:ex_pair}) ($K=1$ solid line, $K=4$ dashed line), while symbols represent the average of 250 serious outbreaks ($K=1$ filled circles, $K=4$ triangles). (a) Homogeneous network. (b) Erd{\H o}s-R{\'e}nyi random graph.}
		\label{fig:smallpox}
	\end{figure}

	The second example we consider is smallpox, a viral disease that has been eradicated globally except for two stocks kept in the secure labs and being used for further research. Several papers have modelled the effectiveness of smallpox when used as a bio-weapon, as well strategies for its containment during possible outbreaks (Ferguson et al. 2003; Kaplan Craft and Wein 2002; Meltzer et al. 2001) \cite{ferguson2003planning,kaplan,metz}. Due to a profound impact smallpox has had on a human population over several centuries, an extensive and quite accurate data has been collected about its transmission. Smallpox is spread through a contact with the mucus of an infected individual, which implies that a close contact is essential for a successful disease transmission. In Fig.~\ref{fig:smallpox} we show the simulations of smallpox outbreaks on homogeneous and Erd{\H o}s-R{\'e}nyi networks using parameter values from Table~\ref{tab:distributions} compared to results of the numerical solution of the corresponding pairwise model (\ref{eq:ex_pair}). The first important observation that the higher severity of epidemics outbreaks as suggested by these data makes the pairwise model more accurate, as expected. The effect of including the realistic distribution of infectious period is more pronounced in this case as compared to the SARS simulations, which can be attributed to the fact that smallpox model includes four stages of infection, while the SARS model had only three stages. Despite changes in the intermediate behaviour for smallpox being more pronounced compared to SARS, the final size of an epidemic as given by the pairwise model only increases from 96.34\% to 97.89\%, which is consistent with an earlier observation that the effect of increasing the number of stages on the final epidemic size is less noticeable for higher $K$.
	
\begin{figure}
		\hspace{-1cm}
		\epsfig{file=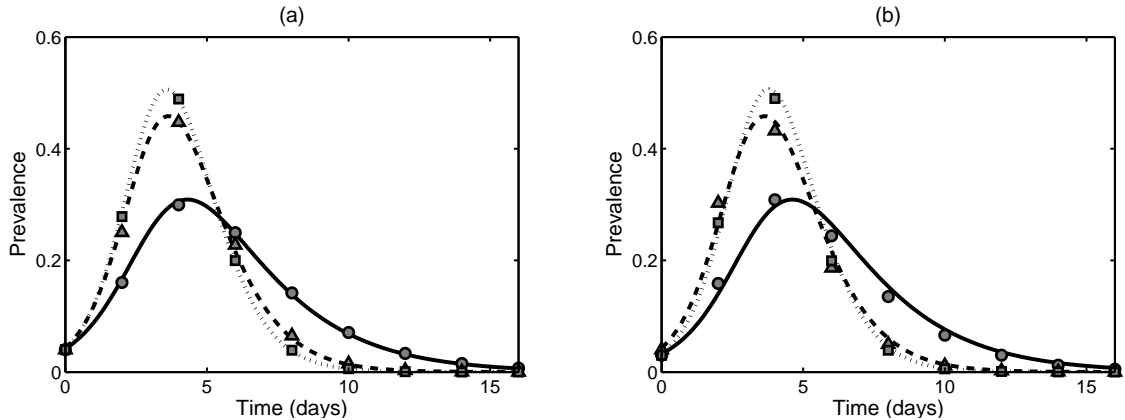,width=18cm}
		\caption{Simulation of an influenza outbreak using data from Table~\ref{tab:distributions} with $n=10$ and $N=1000$. Lines correspond to a numerical solution of the pairwise model (\ref{eq:ex_pair}) ($K=1$ solid line, $K=3$ dashed line, $K=5$ dotted line), while symbols represent the average of 250 serious outbreaks ($K=1$ filled circles, $K=3$ triangles, $K=5$ squares). (a) Homogeneous network. (b) Erd{\H o}s-R{\'e}nyi random graph.}
		\label{fig:flupair}
	\end{figure}

	Figure~\ref{fig:flupair} illustrates the comparison of a pairwise model (\ref{eq:ex_pair}) with the closure (\ref{eq:closure_pair}) and a stochastic simulation on the example of influenza data with different number of stages of infection. Comparison of figures (a) and (b) shows that the heterogeneity introduced by the degree distribution makes the pairwise model less accurate due to the fact that this model only takes into account the mean degree $n$. This suggests that whilst our model is very helpful for understanding general features of multi-stage disease dynamics on networks, it has to be extended further to deal effectively with wider and more realistic node degree distributions. One should note that the effects of increasing the number of stages on peak prevalence and the duration of epidemics
	reduce for higher values of $K$, as can be observed by comparing the minor changes between temporal profiles of the three- and five-stage influenza epidemics presented as shown in Fig.~\ref{fig:flupair}.
	
	\section{Discussion}
	
	In this paper we have analysed the behaviour of multi-stage infections with particular emphasis on contact networks. Unlike the well-mixed models, for which the number of stages modifies the temporary profile of an outbreak but does not affect the final epidemic size or the condition for disease outbreak, in the case of disease spread on a network, the number of stages, i.e. the precise distribution of infectious period, plays a much more prominent role.
	
	In order to make analytical progress with the analysis of disease dynamics on networks, we have employed the framework of pairwise approximation. This has allowed us to determine the probability of disease transmission across a network edge and to find an ${\mathcal R}_0$-like threshold that controls the onset of epidemics. We have also derived an analytical expression for the final size of an epidemic, which is in perfect agreement with the final size computed using percolation theory (Newman 2002; Kenah and Robins 2007) \cite{newmannetworks, kenahrobins}, and therefore, our findings can be considered exact in the limit of infinite population size. All of these quantities depend not only on the basic disease characteristics, such as, the transmission rate and the average infectious period, but also on the distribution of the infectious period as represented by the number of stages in the model. The importance of this result lies in the fact that unlike earlier studies of multi-stage models in well-mixed populations (Anderson and Watson 1980; Ma and Earn 2006) \cite{anderson1980spread,ma2006generality}, for the same average duration of the infection period, the final epidemic size is not constant but increasing with the number of stages. We also observe that the threshold at which point a major epidemic is expected depends on the number of infectious stages, with epidemics becoming more likely as the number of stages is increased. This dependence emerges due to the higher resolution of our model which allows us to identify new links between model ingredients and disease dynamics. Similar results have been noted in related studies, for example, in models concerned with contact tracing (Eames and Keeling 2003) \cite{ekcontact} and models of coupled disease and information transmission on networks (Funk et al. 2009) \cite{funk}.
	
	Numerical simulations of epidemic outbreak for several different multi-stage infections demonstrate that while the pairwise model provides a reasonably good approximation of the network dynamics, the agreement with stochastic simulations is affected by clustering and local network structure that can induce correlations in the dynamics of different nodes, as well as the inhomogeneity in the node degree distribution, as should be expected from the fact that the pairwise closure only depends on the average node degree.
	
	There are several directions in which the approach presented in this paper could be extended. These include the analysis of SIS and SEIR models, as well as inclusion of multiple stages for both the latent and infected classes (Nguyen and Rojani 2008) \cite{nguyen2008noise}. Whilst inclusion of latent classes may have no effect on the basic reproduction number or the final size distribution in a homogeneous model (Black and Ross 2015; House et al. 2013) \cite{black,house13}, whether the same would be true in a network model remains to be seen. Another interesting and important problem would be the consideration of network dynamics for epidemic models with temporary immunity (Blyuss and Kyrychko 2010) \cite{BK10}. Allowing the level of infectiousness of different nodes to vary depending on the stage of infection they belong to would result in even more realistic models of multi-stage diseases on networks. One of the challenging but practically important generalisations of the present framework would be an extension of a pairwise model that would account for heterogeneity in node degree distribution (House and Keeling 2011a) \cite{house2011insights}. This would provide deterministic models potentially amenable to analytical treatment that would more accurately represent stochastic disease dynamics.\\
	
	{\bf Acknowledgements} N. Sherborne acknowledges funding for his PhD studies from the Engineering and Physical Sciences Research Council (EPSRC). The authors would like to thank the anonymous referees for helpful comments and suggestions that have helped to improve the presentation.

	\section*{Appendix A}
	
	In this Appendix we prove an expression (\ref{eq:Kstage}) for the probability of transmission across a given link. For an arbitrary number of stages and transition/recovery parameter $K\gamma$, the distribution of the infectious period is gamma distributed, and hence we consider here the density function originally stated in (\ref{eq:gamma_tau})
	\[
	g(x;K,1/(K\gamma)) = \frac{1}{(K-1)!}(K\gamma)^K x^{K-1} e^{-K\gamma x}.
	\]
	Since the probability of infection taking place for a given $S-I$ link during time $t$ is given by $1-e^{-\tau t}$, the probability of transmission across this link in a $K$-stage is given by
	\begin{equation}\label{eq:prepart}
	\begin{array}{l}
	\displaystyle{\tilde{\tau}= \int_0^{\infty} (1 - e^{-\tau x})\left(\frac{1}{(K-1)!}(K\gamma)^K x^{K-1} e^{-(K\gamma)x}\right)dx}\\\\
	\displaystyle{= \frac{(K\gamma)^K}{(K-1)!}\left[\int_0^{\infty} x^{K-1} e^{-(K\gamma)x}dx - \int_0^{\infty} x^{K-1} e^{-(\tau + K\gamma)x}dx\right]}\\\\
	\displaystyle{= 1 - \frac{(K\gamma)^K}{(K-1)!}\int_0^{\infty} x^{K-1} e^{-(\tau + K\gamma)x}dx,}
	\end{array}
	\end{equation}
	where the final equality is obtained by noting that the first integral is simply the integral of the gamma distribution function over $\mathbb{R}^+$, and, hence, is equal to one. Integration by parts yields a recursive relation
	\[
	\int_0^{\infty}x^{K-1}e^{-(\tau + K\gamma)x}dx=\frac{K-1}{\tau + K\gamma}\int_0^{\infty} x^{K-2} e^{-(\tau + K\gamma)x}dx,
	\]
	which is valid for any integer $K>1$, and this then implies
	\[
	\int_0^{\infty}x^{K-1}e^{-(\tau + K\gamma)x}dx=\frac{(K-1)!}{(\tau + K\gamma)^{K}}.
	\]
	Substituting this expression into Eq. (\ref{eq:prepart}) yields
	\[
	\tilde{\tau}=1 - \frac{(K\gamma)^K}{(K-1)!} \frac{(K-1)!}{(\tau + K\gamma)^K}=1-\frac{(K\gamma)^K}{(\tau + K\gamma)^K}.
	\]

	\section*{Appendix B}
	
	Linearisation of the pairwise model (\ref{eq:ex_pair}) with the closure (\ref{eq:closure_pair}) at the disease-free equilibrium yields the stability condition for eigenvalues $\lambda$ as a $(2K+2) \times (2K+2)$ matrix. It is useful to first consider it in a block form as follows,
	\[
	\left(\begin{array}{cc}
	A & B \\
	C & D \end{array}\right)
	\]
	where $C$ is a zero $(K+1) \times (K+1)$ matrix, and the matrix $A$ is lower-diagonal, and therefore, its determinant is the product of the diagonal terms. Hence, the characteristic equation can be written as
	\[
	\lambda^2(\lambda+K\gamma)^K
	\left|\begin{array}{ccccc}
	\tau(n-1) - K\gamma - \tau - \lambda& \tau(n-1) 								& & \ldots & \tau(n-1) \\
	K\gamma							& -K\gamma - \tau - \lambda & 0 & \ldots & 0			 \\
	0										&	K\gamma										& \ddots & \ddots & \vdots \\
	\vdots							& 0													& \ddots & \ddots & 0			\\
	0										& \ldots										& 0			 & K\gamma & -K\gamma - \tau - \lambda 
	\end{array}\right| =0
	\]	
	This matrix can now be reduced to a series of lower-diagonal matrices to give the following general form of the characteristic equation
	\[
	\begin{array}{l}
	\displaystyle{0= \lambda^2(\lambda + K\gamma)^K\Bigg[(\tau(n-1) - K\gamma - \tau - \lambda)(-K\gamma - \tau - \lambda)^{K-1}}\\\\
	\displaystyle{\left. \hphantom{{}=\tau(n-1)} + \tau(n-1)\left(\sum_{i=1}^{K-1}(-1)^{K-i}(K\gamma)^{K-i}(-K\gamma - \tau - \lambda)^{i-1}\right)\right]}\\\\
	\displaystyle{= \lambda^2(\lambda + K\gamma)^K\Bigg\{(\tau + K\gamma + \lambda)^K}\\
	\displaystyle{\left. \hphantom{{}=\tau}	- \tau(n-1)\left[(\tau + K\gamma + \lambda)^{K-1} + \sum_{i=1}^{K-1}(K\gamma)^{K-i}(\tau + K\gamma + \lambda)^{i-1}\right]\right\}.}
	\end{array}
	\]
	It immediately follows that the above equation has roots of $\lambda=0$, $\lambda=-K\gamma$, and the other $K$ roots are determined by the roots of the expression in curly brackets. Since an epidemic outbreak occurs when the disease-free equilibrium becomes unstable, one has to identify conditions on parameters when the stability of the disease-free steady state changes, i.e. where $\lambda = 0$. Substituting $\lambda = 0$ into the expression in curly brackets yields
	\[
	\begin{array}{l}
	0\displaystyle{= (\tau + K\gamma)^K + \tau(n-1)\left[(\tau + K\gamma)^{K-1} - \sum_{i=1}^{K-1}(K\gamma)^{K-i}(\tau + K\gamma)^{i-1}\right]}\\\\
	\displaystyle{= (\tau + K\gamma)^K - (n-1)\Big[(\tau + K\gamma)^{K} - (K\gamma)^K\Big].}
	\end{array}
	\]
	This relation can be recast as
	\[
	1=(n-1)\left(1 - \frac{(K\gamma)^K}{(\tau + K\gamma)^K}\right)=(n-1)\tilde{\tau},
	\]
	which gives the desired expression of $\mathcal{R}=(n-1)\tilde{\tau}$ in Eq. (\ref{eq:R}).

	\section*{Appendix C}
	
	To prove relation (\ref{eq:SR}), we consider the time derivatives of the functions $P_i = \frac{[SI_i]}{[S]^a}$ for $i=1,2, \ldots, K$, which can be found from the pairwise model (\ref{eq:ex_pair}):
	\[
	\begin{array}{l}
	\dot{P_1}= -(\tau + K\gamma)P_1 +\tau a \frac{[SS]}{[S]} F,\\
	\dot{P_i}= -(\tau + K\gamma)P_i + K\gamma P_{i-1}, \quad i = 2, 3, \ldots, K.
	\end{array}
	\]
	We also remind the reader of the functions $G$ and $L$ and equations for their dynamics
	\[
	\begin{array}{l}
		G = \frac{[SR]}{[S]^a} \Longrightarrow \dot{G} = K\gamma P_K ,\
		L = \frac{[SS]}{\exp\left(n[S]^{1/n}\right)[S]^{2a}} \Longrightarrow \dot{L} = -a \tau \frac{[SS]}{[S]}F.
	\end{array}
	\]
	Integrating the equation for $P_1$ and using the fact that $[SI_1](0)=[SI_1](\infty)=0$, gives
	\begin{equation}\label{eq:P1}
	\begin{array}{l}
	\displaystyle{0 = \int_0^{\infty} \dot{P_1} dt = -(\tau + K\gamma)\int_0^{\infty}P_1 dt + a\tau\int_0^{\infty} \frac{[SS]}{[S]}F dt}\\\\
	\displaystyle{= -(\tau + K\gamma)\int_0^{\infty}P_1 dt - [L(\infty) - L(0)].}
	\end{array}
	\end{equation}
	In a similar way, integrating the equation for $P_2$ yields
	\[
	0=\int_0^{\infty} \dot{P_2} dt = -(\tau + K\gamma)\int_0^{\infty} P_2 dt + K\gamma\int_0^{\infty} P_1 dt,
	\]
	which can be rewritten as
	\[
	\int_0^{\infty} P_1 dt = \frac{\tau +K\gamma}{K\gamma}\int_0^{\infty} P_2 dt.
	\]
	Proceeding the the same way, one obtains
	\[
	\int_0^{\infty} P_i dt = \frac{\tau +K\gamma}{K\gamma}\int_0^{\infty} P_{i+1} dt,\quad i=2,3,\ldots,K-1.
	\]
	Going through all stages of infections, we find
	\[
	\displaystyle{\int_0^{\infty} P_1 dt = \frac{(\tau +K\gamma)^{K-1}}{(K\gamma)^{K-1}}\int_0^{\infty} P_K dt.}
	\]
	On the other hand, integrating equation for $G$ and using $G(0)=0$ gives
	\[
	\displaystyle{G(\infty)-G(0)=\frac{[SR]_{\infty}}{[S]_{\infty}^a}=K\gamma\int_0^{\infty} P_K dt \quad\Longrightarrow\quad \int_0^{\infty} P_K dt= \frac{1}{K\gamma}\frac{[SR]_{\infty}}{[S]_{\infty}^a}.}
	\]
	Combining the last two expressions, we obtain
	\[
	\int_0^{\infty} P_1 dt = \frac{(\tau + K\gamma)^{K-1}}{(K\gamma)^K}\frac{[SR]_{\infty}}{[S]_{\infty}^a},
	\]
	and substituting this result into Eq. (\ref{eq:P1}) gives the final relation (\ref{eq:SR}):
	\begin{equation}
	\frac{[SR]_{\infty}}{[S]_{\infty}^a} = (\tilde{\tau}-1)[L(\infty) - L(0)].
	\end{equation}
	
	In order to prove relation (\ref{eq:SS}), we examine the ratio $[SS]/[S]$, whose dynamics is governed by the following equation
	\[
	\frac{d}{dt}\frac{[SS]}{[S]}= -\tau\frac{(n-2)}{n}\frac{[SS]}{[S]}\frac{\sum_{i=1}^K[SI_i]}{[S]}.
	\]
	Separating variables and integrating this equation gives
	\begin{equation}\label{eq:SS/S}
	\left[\ln\left(\frac{[SS]}{[S]}\right)\right]_0^{\infty}= -\tau \frac{(n-2)}{n}\int_0^{\infty} \frac{\sum_{i=1}^K[SI_i]}{[S]} dt.
	\end{equation}
	Rather than compute the integral in the right-hand side of the above equation, we use the first equation of the pairwise model (\ref{eq:ex_pair}), which can be written as
	\[
	\frac{1}{[S]}\frac{d}{dt}[S]= -\tau\frac{\sum_{i=1}^K[SI_i]}{[S]}.
	\]
	Integrating this equation gives
	\[
	\int_0^{\infty} \frac{1}{[S]} d[S]= -\tau\int_0^{\infty}\frac{\sum_{i=1}^K[SI_i]}{[S]} dt \quad\Longrightarrow\quad
	\left(\ln[S]\right)_0^{\infty}= -\tau\int_0^{\infty}\frac{\sum_{i=1}^K[SI_i]}{[S]} dt.
	\]
	Using this expression to replace an integral in (\ref{eq:SS/S}) gives
	\[
	\ln\left(\frac{[SS]_{\infty}}{[S]_{\infty}}\right)- \ln\left(\frac{[SS]_{0}}{[S]_{0}}\right)= \frac{n-2}{n}\ln\left(\frac{[S]_{\infty}}{[S]_0}\right).
	\]
	Substituting $[S]_0=N$ and $[SS]_0=nN$, this formula can be rewritten as
	\[
	\ln\left(\frac{[SS]_{\infty}}{[S]_{\infty}}\right)= \ln\left(\frac{nN}{N}\right)+\ln\left(\frac{[S]_{\infty}}{N}\right)^{\frac{n-2}{n}},
	\]
	or alternatively,
	\[
	\frac{[SS]_{\infty}}{[S]_{\infty}}= n\left(\frac{[S]_{\infty}}{N}\right)^{\frac{n-2}{n}}.
	\]
	Multiplying both sides by $[S]_{\infty}$ and using the definition $a=(n-1)/n$, we obtain
	\[
	[SS]_{\infty}=n\frac{[S]_{\infty}^{2(n-1)/n}}{N^{(n-2)/n}}=n\frac{[S]_{\infty}^{2a}}{N^{a-1/n}},
	\]
	which gives the desired relation (\ref{eq:SS}).
	
\end{document}